\documentclass{article}

\usepackage{arxiv}

\usepackage[utf8]{inputenc}
\usepackage[T1]{fontenc}
\usepackage{amsmath,amssymb,amsfonts}
\usepackage{graphicx}
\usepackage{booktabs}
\usepackage{multirow}
\usepackage{array}
\usepackage{url}
\usepackage{microtype}
\usepackage[hidelinks]{hyperref}
\usepackage{float}
\usepackage{placeins}
\graphicspath{{./figs/}}

\title{Not All Agents Are Equal: Code Quality and Post-Merge Maintenance Across Five Autonomous Coding Agents in the Wild}

\author{
 Obada Kraishan \\
  College of Media and Communication\\
  Texas Tech University\\
  Lubbock, TX 79409, USA \\
}

\begin{document}
\maketitle

\begin{abstract}
Autonomous coding agents now open pull requests in public repositories at a scale that was out of reach two years ago, yet little is known about what happens to that code after it lands. This paper studies 37,623 provenance-labeled pull requests (PRs) from five commercial agents (OpenAI Codex, Devin, GitHub Copilot, Cursor, and Claude Code) and a matched human baseline, drawn from 2,807 GitHub repositories between December 2024 and July 2025. We combine the AIDev dataset with 58{,}792 cached GitHub API responses to measure security smells in added code, structural maintainability, post-merge churn, revert rates, and human review behavior. Three results stand out. First, quality differences are vendor-specific rather than uniform: Codex-authored PRs were reverted about half as often as human PRs (6.1\% vs.\ 11.5\%, odds ratio 0.50), while Devin PRs were reverted more often (14.5\%, odds ratio 1.31). Second, agent code pooled across vendors was less likely than human code to contain a security smell (odds ratio 0.63), driven by fewer hardcoded credentials and \texttt{eval}-style constructs. Third, review effort concentrates unevenly: Copilot PRs drew the most human reviews and change requests, and Claude Code PRs waited the longest for a first human review (median 12.6 hours). All pipeline code, statistical reports, and figures are released for replication.
\end{abstract}


\section{Introduction}
Software teams have moved quickly from code completion to delegation. Coding agents built on large language models no longer suggest single lines; they take an issue, edit a repository, run tests, and open a pull request under their own name [1]. The AIDev dataset alone records close to a million such agentic PRs across more than one hundred thousand repositories in under a year [1]. Adoption at this pace raises a practical question for the projects that receive this code: what does it cost to keep it?

Most evidence on AI-generated code quality comes from controlled settings. Benchmark studies measure whether models solve isolated tasks [2], [3], agent frameworks are compared on curated issue sets [4], [5], and security work has largely prompted models in the lab and inspected the output [7]--[9]. These designs answer what models \emph{can} produce. They say less about what agents \emph{do} produce inside real projects, where code must pass review, coexist with an existing design, and survive the months after merge. Field evidence is thinner and mostly limited to one assistant, Copilot, in its autocomplete form [10]--[12].

This paper examines agent-authored code where it actually lives. We study 37,623 PRs with known authorship, of which 33,596 were opened by five commercial agents and 4,027 by humans in the same repositories over the same window, and we follow each merged change for 90 days. The comparison covers five dimensions: defect-prone patterns in the added code, structural maintainability, post-merge churn, revert rates, and the human review process around each PR. Because every PR carries a vendor label, the analysis can separate ``agents in general'' from the behavior of individual products, a distinction that turns out to matter.

We organize the study around five research questions.
\textbf{RQ1:} Do agent-authored PRs carry a higher density of defect-prone code patterns than human PRs?
\textbf{RQ2:} Which agents produce the most maintainable code?
\textbf{RQ3:} What post-merge maintenance overhead does agent code impose?
\textbf{RQ4:} Are particular vulnerability classes over-represented in agent code?
\textbf{RQ5:} How does human review behavior differ across agents?

The paper makes three contributions. (i) A reproducible measurement pipeline that joins AIDev provenance labels with GitHub enrichment, patch-level static analysis over 1.35 million added lines, and a 90-day maintenance window for 26,283 merged PRs. (ii) Statistical evidence, with effect sizes and multiplicity control, that code quality and maintenance outcomes differ more \emph{between} agents than between agents and humans on several measures. (iii) A released artifact (pipeline code, a 63-variable codebook, per-question statistical reports, and figures) that supports replication and extension as new agents appear.

The rest of the paper is organized as follows. Section~2 reviews related work. Section~3 describes the corpus, enrichment, metrics, and statistical design. Section~4 reports results by research question. Section~5 discusses implications, Section~6 examines threats to validity, and Section~7 concludes.

\section{Related Work}
This section positions the study against three lines of work: evaluation of code-generation models, security analysis of AI-produced code, and repository mining of AI-assisted development.

\subsection{Evaluating code-generating models and agents}
Benchmarks such as HumanEval established functional correctness as the standard yardstick for code models [2]. SWE-bench moved evaluation toward realistic repository issues [3], and agent scaffolds such as SWE-agent and OpenHands compete on that benchmark family [4], [5]. Cemri et al.\ [6] analyzed why multi-agent LLM systems fail, using annotated execution traces rather than repository outcomes. All of these efforts score task success in controlled runs; none observes how accepted agent code behaves inside living projects afterward, which is the gap this study addresses.

\subsection{Security of AI-generated code}
Pearce et al.\ [7] found that roughly 40\% of Copilot completions in security-relevant scenarios were vulnerable. Perry et al.\ [8] and Sandoval et al.\ [9] reached partly conflicting conclusions in user studies of AI-assisted programmers, and Fu et al.\ [10] audited Copilot-generated snippets found in public projects. These studies mostly analyze suggested or freshly generated code. Our setting differs in two ways: the unit is a full PR written autonomously by an agent, and the comparison group is human PRs in the same repositories rather than synthetic prompts.

\subsection{Mining AI-assisted development at scale}
Ziegler et al.\ [11] and Peng et al.\ [12] measured productivity effects of Copilot on its users. An industry analysis of version-control history argued that AI assistance coincides with rising code churn [13]. The AIDev dataset [1] made agent-authored PRs observable at population scale with per-vendor labels, and its authors characterize adoption patterns and PR outcomes. We build directly on AIDev and extend it with follow-up maintenance signals, patch-level quality measures, and a matched human baseline, while following established guidance on the pitfalls of mining GitHub [14]. Classic work on code review [15] informs our RQ5 measures.

\section{Data and Method}
\begin{figure}[t]
\centering
\includegraphics[width=\textwidth]{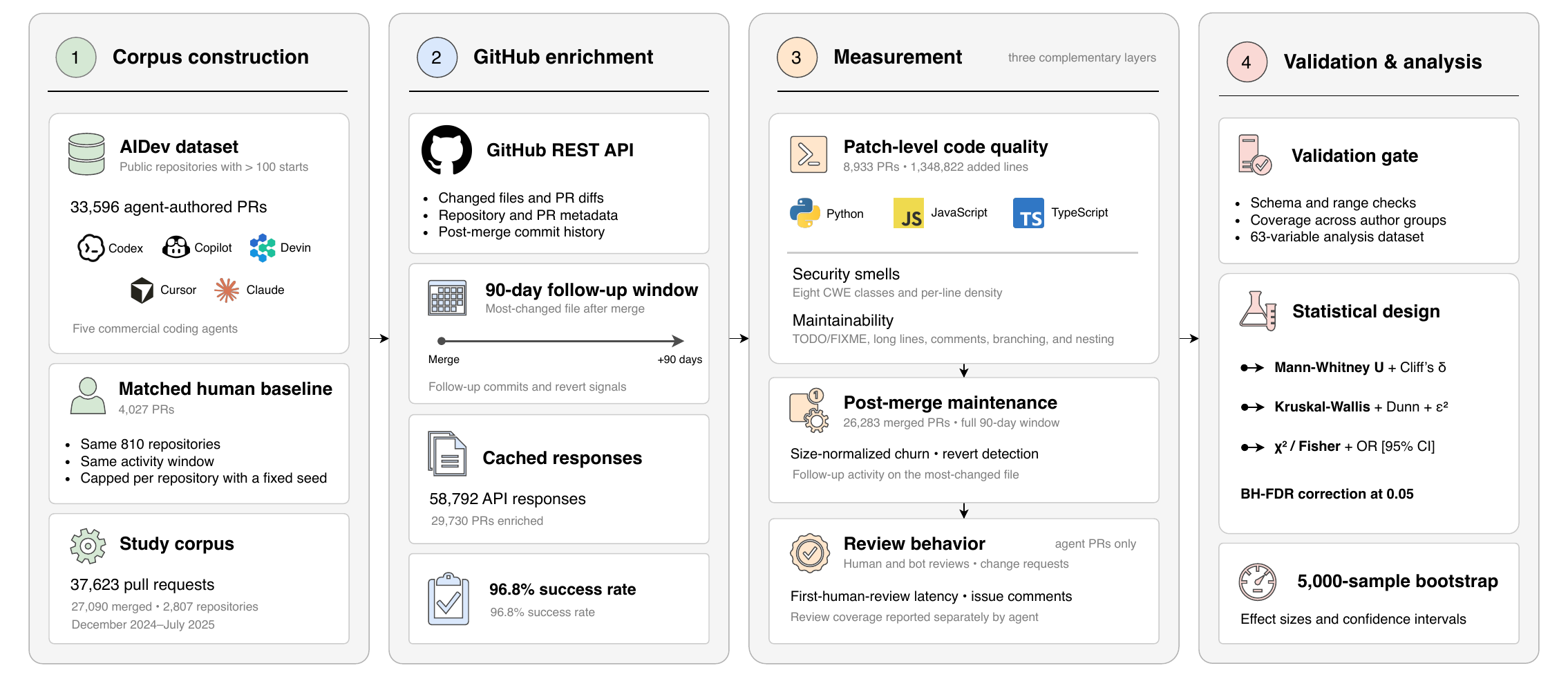}
\caption{Overview of the study pipeline: corpus construction from AIDev with a matched human baseline, GitHub enrichment with a 90-day follow-up window, patch-level and post-merge measurement, and the validation and statistical-analysis stage. Agent and language marks identify the five products and the three analyzed languages.}
\label{fig:pipeline}
\end{figure}

This section describes how the corpus was assembled, how each measure is computed, and how the statistical analysis is designed. Fig.~\ref{fig:pipeline} gives an overview of the four stages, all of which are implemented in the released repository.

\subsection{Corpus construction}
The starting point is the curated slice of AIDev [1], which covers agentic PRs in repositories with more than 100 stars. We downloaded seven normalized tables, including 33,596 agent PRs labeled OpenAI Codex, Devin, GitHub Copilot, Cursor, or Claude Code, and AIDev's human PR sample drawn from the same repositories. The human sample was tightened into a matched baseline: we kept human PRs only from repositories that also contain agent PRs (810 repositories), restricted them to the agents' activity window (December 24, 2024 to July 30, 2025), and capped the number per repository under a fixed random seed so that no single project dominates. The result is 4,027 human PRs alongside 9,750 agent PRs in the same 810 repositories, with the remaining agent PRs providing additional within-agent power. Table~\ref{tab:corpus} summarizes the corpus.

\begin{table}[t]
\caption{Corpus composition. Enrich.\ = PRs with successful GitHub enrichment; Anal.\ = PRs whose diffs contained Python, JavaScript, or TypeScript lines and entered static analysis; Med.\ = median changed lines.}
\label{tab:corpus}
\centering
\small
\begin{tabular}{lrrrrr}
\toprule
Group & PRs & Merged & Enrich. & Anal. & Med. \\
\midrule
OpenAI Codex   & 21{,}799 & 18{,}004 & 17{,}761 & 4{,}570 & 63 \\
Devin          &  4{,}827 &  2{,}595 &  2{,}186 & 1{,}224 & 61 \\
GitHub Copilot &  4{,}970 &  2{,}139 &  3{,}459 & 1{,}102 & 76 \\
Cursor         &  1{,}541 &  1{,}005 &    946   &   571   & 96 \\
Claude Code    &    459   &    271   &    456   &   241   & 495 \\
Human          &  4{,}027 &  3{,}076 &  3{,}038 & 1{,}225 & 52 \\
\midrule
Total          & 37{,}623 & 27{,}090 & 27{,}846 & 8{,}933 & \phantom{0} \\
\bottomrule
\end{tabular}
\end{table}

\subsection{GitHub enrichment}
AIDev records PR metadata but not post-merge history. For every PR we queried the GitHub REST API for the list of changed files and, for merged PRs, for commits touching the PR's most-changed file during the 90 days after merge. The follow-up query also flags revert-style commits by message. All 58,792 API responses were cached to disk, which makes the collection resumable and lets later stages reread raw payloads without further requests. Of 30,721 PRs submitted to enrichment, 29,730 (96.8\%) succeeded; 991 (3.2\%) referred to repositories or PRs that had been deleted or legally blocked since AIDev's snapshot, an attrition rate we report rather than impute.

\subsection{Patch-level static analysis}
Quality measures are computed on the \emph{added} lines of each PR's diff, read directly from the cached file-level API payloads. We analyze the three most common languages in the corpus, namely Python, JavaScript, and TypeScript, which yields 8,933 PRs and 1{,}348{,}822 added lines (Table~\ref{tab:corpus}, Anal.). Diffs above 5,000 added lines are excluded as auto-generated. Two families of measures are extracted. Maintainability counts cover TODO/FIXME-style markers, lines longer than 120 characters, comment lines, branching keywords, and maximum indentation depth. Security smells are matched by a regex catalog mapped to eight Common Weakness Enumeration (CWE) classes [19]: hardcoded credentials (CWE-798), SQL string construction in query calls (CWE-89), shell command injection (CWE-78), \texttt{eval}/\texttt{exec} use (CWE-95), unsafe deserialization (CWE-502), weak hash or cipher choices (CWE-327), DOM-based XSS sinks (CWE-79), and concatenated paths in file-open calls (CWE-22). Pattern matching on diff lines cannot confirm exploitability; we therefore interpret hits as \emph{smells}, in line with prior corpus studies [7], [10], and return to the limitation in Section~6.

For PR $i$ with $L_i$ analyzed added lines, $S_i$ security hits, and $M_i$ maintainability hits, the two density measures are
\begin{equation}
s_i = 100\,\frac{S_i}{L_i}, \qquad m_i = 100\,\frac{M_i}{L_i}.
\label{eq:density}
\end{equation}
For merged PR $i$ with $C_i$ changed lines and $F_i$ follow-up commits on its most-changed file within 90 days, churn is
\begin{equation}
c_i = 100\,\frac{F_i}{C_i}.
\label{eq:churn}
\end{equation}

\subsection{Review measures}
AIDev ships review, comment, and timeline tables for agentic PRs. Reviews are split by the platform's account type; bot reviews (continuous-integration and review bots) are counted separately, and latency and depth are computed from human reviewers only. Review latency is the time from PR creation to the first human review. These tables do not cover the human PR sample, so RQ5 compares agents with one another and reports per-group coverage explicitly.

\subsection{Statistical design}
Software metrics of this kind are heavy-tailed and zero-inflated, so the analysis is non-parametric throughout. Group differences use the Mann--Whitney $U$ test with Cliff's $\delta$ as effect size,
\begin{equation}
\delta = \frac{2U}{n_1 n_2} - 1,
\label{eq:delta}
\end{equation}
interpreted with the customary bands ($|\delta|<0.147$ negligible, $<0.33$ small, $<0.474$ medium, otherwise large) [16]. Omnibus differences across the six author groups use the Kruskal--Wallis $H$ test with
\begin{equation}
\varepsilon^{2} = \frac{H - k + 1}{n - k}
\label{eq:eps}
\end{equation}
as effect size [18], followed by Dunn's test for pairwise contrasts. Binary outcomes (revert, presence of a CWE class) use $\chi^2$ or Fisher's exact test with odds ratios and 95\% confidence intervals. Every family of related comparisons is corrected with the Benjamini--Hochberg procedure at a false discovery rate of 0.05 [17]; we report adjusted $p$-values. Confidence intervals for $\delta$ come from a 5,000-resample bootstrap under a fixed seed. A validation gate checks schema, ranges, and group coverage before any analysis runs, and a generated codebook documents all 63 variables.
\section{Results}
This section reports the findings for each research question in turn. Throughout, $p$-values are Benjamini--Hochberg adjusted within their comparison family, and $\delta>0$ means the agent group is stochastically larger than the human group.

\subsection{RQ1: Defect-prone patterns in added code}
Security-smell density (\ref{eq:density}) differs across the six author groups ($H(5)=52.5$, $p<.001$, $\varepsilon^2=.005$), but the pairwise contrasts against humans are small in absolute terms. Codex ($\delta=-.02$, $p=.001$), Copilot ($\delta=-.02$, $p=.025$), and Cursor ($\delta=-.03$, $p=.004$) sit slightly below the human level, Devin is indistinguishable from it ($p=.60$), and Claude Code sits above it ($\delta=+.05$, $p=.004$). All five effects are negligible by the conventional bands, so the honest summary is that per-line smell density is similar across authorship once a PR reaches a public repository.

\begin{figure}[t]
\centering
\includegraphics[width=0.62\textwidth]{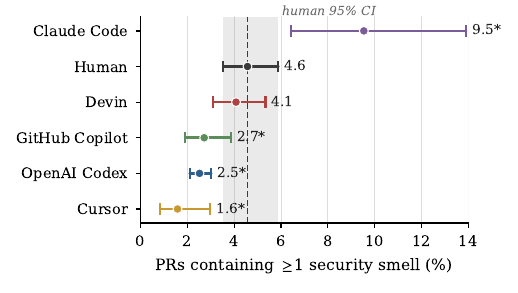}
\caption{Share of analyzed PRs containing at least one security smell, by author group. The dashed line and shaded band mark the human rate and its 95\% confidence interval.}
\label{fig:prev}
\end{figure}

Presence rates tell a sharper story. A PR from the pooled agents is less likely than a human PR to contain any security smell at all: 2.9\% vs.\ 4.6\% ($\chi^2(1)=8.59$, $p=.003$, odds ratio $0.63$, 95\% CI $[0.47, 0.85]$). Fig.~\ref{fig:prev} breaks the rate out by group. Cursor (1.6\%) and Codex (2.5\%) are the cleanest; Claude Code is the outlier at 9.5\%, roughly twice the human rate; we return to this point in RQ4 and Section~5. Maintainability-smell density separates groups more strongly ($H(5)=467.7$, $p<.001$, $\varepsilon^2=.052$): Codex adds markedly fewer TODO markers and over-long lines than humans ($\delta=-.19$, small), Devin somewhat fewer ($\delta=-.07$), and the remaining agents are at the human level. A size-stratified sensitivity check found the pooled agent--human difference in security density concentrated in the largest PRs (XL bucket, $\delta=-.08$, $p=.025$), with no difference in the four smaller buckets.

\subsection{RQ2: Maintainability across agents}
All four structural measures differ across groups (nesting depth $H(5)=118.3$, $\varepsilon^2=.013$; branch density $H(5)=104.0$, $\varepsilon^2=.011$; comment ratio $H(5)=868.5$, $\varepsilon^2=.097$; maintainability-smell density $H(5)=467.7$, $\varepsilon^2=.052$; all $p<.001$). Comment ratio is the widest gap: at the median, Copilot (.065), Claude Code (.076), and Cursor (.042) comment their added code, while the median Codex, Devin, and human diff contains no comment lines at all. Claude Code writes the structurally heaviest code (median nesting 4 levels vs.\ 3 elsewhere; highest branch density, .063 per line), consistent with its much larger PRs (median 495 changed lines, Table~\ref{tab:corpus}). Dunn contrasts confirm the pattern: 44 of 60 pairwise tests reach adjusted $p<.05$, and every agent differs from at least two others on at least one measure. No single agent dominates all four dimensions, which argues against treating ``agent code'' as one population.

\subsection{RQ3: Post-merge maintenance overhead}
The maintenance analysis covers 26,283 merged PRs with a full 90-day window. On raw follow-up commits, Copilot-touched files see the least subsequent activity of any group, humans included ($\delta=-.20$ vs.\ humans, small, $p<.001$); Cursor and Claude Code are also below the human level (both negligible), and Codex slightly above it ($\delta=+.03$, $p=.020$). Normalizing by PR size (\ref{eq:churn}) sharpens the ranking: Claude Code shows the lowest churn per changed line of any group ($\delta=-.33$ vs.\ humans, medium, $p<.001$; median $c_i=0.5$ vs.\ $5.3$ for humans), followed by Copilot ($\delta=-.16$, small) and Cursor ($\delta=-.08$), with Codex and Devin at the human level.

\begin{table}[t]
\caption{Revert outcomes within 90 days of merge, each agent vs.\ the human baseline (11.5\%). OR $<1$ means the agent's PRs were reverted less often. $p$ values are BH-adjusted.}
\label{tab:reverts}
\centering
\small
\begin{tabular}{lrrcc}
\toprule
Agent & $n$ & Reverted & OR [95\% CI] & $p$ \\
\midrule
OpenAI Codex   & 17{,}756 &  6.1\% & 0.50 [0.44, 0.57] & $<.001$ \\
Devin          &  2{,}185 & 14.5\% & 1.31 [1.11, 1.54] & .004 \\
GitHub Copilot &  2{,}094 & 12.5\% & 1.10 [0.93, 1.31] & .457 \\
Cursor         &    946   & 11.4\% & 1.00 [0.79, 1.25] & 1.00 \\
Claude Code    &    267   & 10.5\% & 0.90 [0.60, 1.36] & .878 \\
\bottomrule
\end{tabular}
\end{table}

\begin{figure}[t]
\centering
\includegraphics[width=0.62\textwidth]{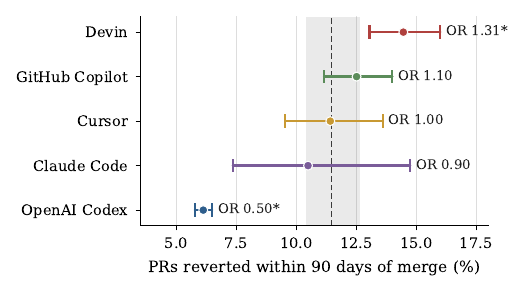}
\caption{Ninety-day revert rate by agent, with 95\% confidence intervals. The dashed line and shaded band mark the human baseline and its confidence interval; asterisks mark BH-adjusted $p<.05$.}
\label{fig:rev}
\end{figure}

Reverts, the costliest failure mode, split the vendors cleanly (Table~\ref{tab:reverts}, Fig.~\ref{fig:rev}). Codex PRs were reverted at about half the human rate, Devin PRs at about 1.3 times it, and the other three agents were statistically indistinguishable from humans. Reading the churn and revert results together: none of the five agents imposes a measurably higher per-line maintenance burden than human contributors in the same repositories, and two of them differ from humans in opposite directions on the most severe outcome.

\subsection{RQ4: Vulnerability classes}
\begin{figure}[t]
\centering
\includegraphics[width=0.62\textwidth]{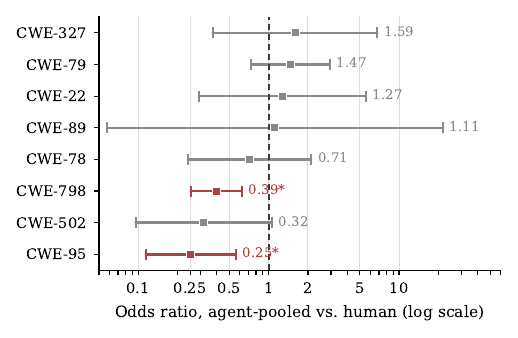}
\caption{Odds ratios (agent-pooled vs.\ human) for the presence of each CWE class per PR, with 95\% CIs; values left of 1 mean the class is rarer in agent code. Red marks BH-adjusted $p<.05$.}
\label{fig:forest}
\end{figure}

Fig.~\ref{fig:forest} shows, for each CWE class, the odds that an agent PR contains the class relative to a human PR. Two classes are under-represented in agent code after correction: hardcoded credentials (CWE-798; 0.9\% of agent PRs vs.\ 2.2\% of human PRs; OR $0.39$ [0.25, 0.62], $p<.001$) and \texttt{eval}/\texttt{exec} injection (CWE-95; 0.2\% vs.\ 0.8\%; OR $0.25$ [0.11, 0.56], $p=.006$). No class is over-represented in the pooled agent code; the classes that lean that way (weak crypto, XSS) do not survive correction and rest on small counts. The per-agent view again shows heterogeneity: Claude Code carries the highest rate of hardcoded-credential smells (3.3\% of its analyzed PRs, above the human 2.2\%), while Codex (0.5\%) and Cursor (0.5\%) carry the lowest, a fourfold spread between products on the same weakness class.

\subsection{RQ5: Human review behavior}
\begin{figure}[t]
\centering
\includegraphics[width=0.62\textwidth]{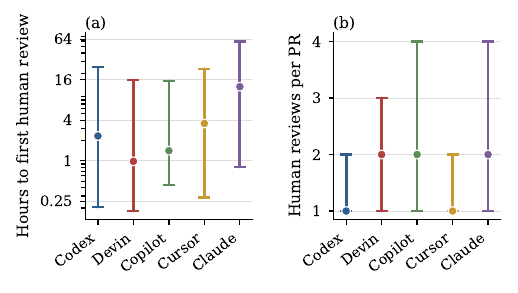}
\caption{Median hours to first human review (left) and mean human reviews per PR (right), agents only; bot reviews excluded. Whiskers span the interquartile range.}
\label{fig:review}
\end{figure}

Review records cover agent PRs only (Section~3.4), with coverage from 5.4\% (Codex) to 51.2\% (Copilot) of each group's PRs, so this question compares agents with one another. All five measures differ across agents (each $p<.001$): hours to first human review ($H(4)=88.9$, $\varepsilon^2=.014$), human reviews per PR ($H(4)=700.6$, $\varepsilon^2=.116$), change requests ($H(4)=293.0$, $\varepsilon^2=.048$), issue comments ($H(4)=243.0$, $\varepsilon^2=.049$), and bot reviews ($H(4)=541.0$, $\varepsilon^2=.119$). Fig.~\ref{fig:review} summarizes the two headline measures. Copilot PRs attract the deepest scrutiny, with 3.6 human reviews and 0.43 change requests per PR on average plus the most bot reviews (4.0), which fits its position as the agent most tightly integrated into the platform's review surface. Claude Code PRs wait longest for a first human review (median 12.6 hours vs.\ 1--4 hours elsewhere), plausibly because its PRs are an order of magnitude larger. Devin and Codex PRs are typically dispatched with one to two quick human reviews.

\section{Discussion}
This section draws out what the numbers mean for the projects receiving agent code and for research on agentic software engineering.

\textbf{Vendor identity is the variable that matters.} On reverts, the most expensive outcome we measure, the gap between the best and worst agent (odds ratios 0.50 vs.\ 1.31 against the same baseline) is far wider than any pooled agent--human gap. The same holds for credential smells (fourfold spread across agents) and review latency. Studies and procurement decisions that treat ``AI-generated code'' as a single category will average away exactly the differences a maintainer cares about.

\textbf{The feared debt has not materialized on these measures within this window.} Agent PRs in popular repositories carry security smells no more often than human PRs, need no more per-line follow-up work, and in several cases need less. Two mechanisms plausibly contribute, and our design cannot separate them: agents may write conservative, template-like code, and the humans steering them may assign bounded, well-specified tasks. Either way, the outcome visible in the repositories is not the churn explosion that early commentary predicted [13], at least not within 90 days of merge in $>$100-star projects.

\textbf{Size is the confound to watch.} Claude Code's higher smell prevalence, heavier structure, and slower first review all co-occur with PRs about eight times the median size of the other groups. Per-line densities and the size-stratified check limit, but cannot remove, this entanglement; per-agent task-mix data would be needed to close it.

\textbf{Review capacity is becoming the bottleneck.} Copilot PRs already consume several human reviews each, and bot reviews outnumber human ones for some agents. As agent volume grows, the scarce resource in this loop is human attention, which argues for research on routing and prioritizing agent PRs rather than only on generating them.

\section{Threats to Validity}
This section lists the main limitations and what we did about them.

\textbf{Construct validity.} Regex smell detection on diff lines flags patterns, not confirmed vulnerabilities, and misses context-dependent flaws; counts should be read as an upper bound on pattern presence, not exploitability. We mitigated this by validating the catalog on planted-defect patches and by correcting one detector during development: an early path-traversal pattern matched relative-import strings, inflating that class roughly 150-fold, and was tightened to fire only inside file-open calls. Churn on the most-changed file is a proxy for whole-PR churn chosen to keep API cost linear; revert detection by commit message misses silent rewrites.

\textbf{Internal validity.} Agents are not randomly assigned to tasks. Repositories and users self-select which agent to use and for what, so group differences mix code quality with task mix. The shared-repository, shared-window human baseline and the size-bucket sensitivity analysis reduce, but do not eliminate, this confounding.

\textbf{External validity.} The corpus covers public repositories with more than 100 stars, three programming languages, and a December 2024--July 2025 window in which Codex contributes 58\% of agent PRs and Claude Code only 459 PRs. Conclusions may not transfer to private codebases, other languages, or later agent versions; the released pipeline is designed to be re-run as the ecosystem shifts.

\textbf{Reliability.} Static coverage (23.7\% of PRs) and review coverage (agents only, 5--51\% by group) are reported per group, and every analysis script regenerates its tables and figures deterministically from a fixed seed.

\section{Conclusion}
This paper followed 37,623 provenance-labeled pull requests from five commercial coding agents and matched human contributors through review, merge, and 90 days of maintenance. Measured in the repositories rather than the lab, agent code is not the uniform liability it is sometimes assumed to be: pooled agents ship fewer security smells than humans, none of the five imposes a higher per-line maintenance burden, and the largest quality gaps run between vendors: Codex reverted at half the human rate, Devin at 1.3 times it. The practical reading is that ``which agent'' is now a measurable engineering decision, and the released pipeline gives maintainers and researchers a way to keep measuring it as the products evolve. Future work should add per-task context to separate agent capability from task assignment, extend the window beyond 90 days, and track whether review capacity keeps pace with agent volume.

\end{document}